\documentclass[%
reprint,
 superscriptaddress,
showpacs,
 amsmath,amssymb,
 aps,
]{revtex4-1}

\usepackage{graphicx}
\usepackage{dcolumn}
\usepackage{bm}
\usepackage{color}
\usepackage[usenames,dvipsnames]{xcolor}

\usepackage[normalem]{ulem}

\begin{document}

\preprint{AIP/123-QED}

\title{Normal Auger processes with ultrashort x-ray pulses in Neon}

\author{Raymond Sullivan}%
\affiliation{Western Illinois University, Macomb, IL 61455}
\affiliation{Argonne National Laboratory, Lemont, IL 60439}
\author{Junteng Jia}%
\affiliation{Department of Chemistry and Chemical Biology, Cornell University, Ithaca, NY 14853}
\affiliation{Argonne Leadership Computing Facility, Argonne National Laboratory, Lemont, IL 60439}
\author{\'Alvaro V\'azquez-Mayagoitia}%
\affiliation{Argonne Leadership Computing Facility, Argonne National Laboratory, Lemont, IL 60439}
\author{Antonio Pic\'on}%
\affiliation{Argonne National Laboratory, Lemont, IL 60439}

\date{\today}

\begin{abstract}
Modern x-ray sources enable the production of coherent x-ray pulses with a pulse duration in the same order as the characteristic lifetimes of core-hole states of atoms and molecules. These pulses enable the manipulation of the core-hole population during Auger decay processes, modifying the lineshape of the electron spectra. In this work, we present a theoretical model to study those effects in Neon. We identify effects in the Auger-photoelectron coincidence spectrum due to the duration and intensity of the pulses. The normal Auger line-shape is recovered in Auger electron spectrum integrated over all photoelectron energies.
\end{abstract}

\maketitle

\section{Introduction}

In recent years, the advent of novel laser-based and accelerator-based sources have enabled the generation of very short (few femtoseconds) x-ray pulses \cite{Allaria2010,Amann2012,Hartmann2014,Ratner2015,Popmintchev2012,Silva2015,Bostedt2016}. The absorption of an x-ray photon has a universal response in any physical system - it predominately produces a core-hole state. Core-hole states usually decay through electron relaxation transitions, such as fluorescence and Auger processes, with lifetimes of hundreds of attoseconds to a few femtoseconds. Interestingly, the aforementioned x-ray sources produce pulses with pulse-lengths that can be shorter than, or comparable to, the characteristic lifetime of core-hole states.  

Short x-ray pulses enable promising applications in time-resolved experiments. In a common pump-probe experiment, the pump pulse excites the system while the probe pulse, delayed by a controlled specific time, interacts with the system and allows to track the pump-induced dynamics. Short x-ray pulses permit the extension of time-resolved x-ray studies into the femtosecond regime and the possibility to study core-hole ultrafast dynamics such as fundamental electron relaxations processes in atoms, molecules, clusters, and nanoparticles. For example, few-femtosecond time-resolved experiments are available to study x-ray induced phenomena by using two ultrashort ($\sim$8 fs) x-ray pulses.\cite{Marinelli2015,Lutman2013} Here, attosecond and femtosecond x-ray pulses may provide additional information about the induced dynamics, as has recently been demonstrated in a few experiments at the Linac Coherent Light Source at SLAC for molecules and clusters. \cite{Schmaltz2015,Picon2016,Lehmann2016,Ken2016} It is then necessary to develop theoretical approaches to explore x-ray interactions with matter at timescales comparable to the core-hole lifetimes.


Short x-ray pulses also enable a particular control of Auger processes as one can tailor the population of the core-hole states while Auger decay occurs. That control causes a fingerprint in the Auger spectrum. Previous studies have demonstrated how such short pulses may affect resonant Auger spectra with a single high-intensity x-ray pulse \cite{Nina2008,Kanter2011,JiCai2010,Nikolopoulos2011,Demekhin2013,Muller2015}, a single x-ray pulse with an optical field, \cite{Mazza2012,Picon2013_A,Picon2013_B,Bernhard2013,Chatterjee2015} and two-color x-ray pulses.\cite{Haxton2014,Picon2015}

The normal (nonresonant) Auger process is usually described by a two step model in which core-hole ionization and Auger decay are treated independently.\cite{Mehlhorn1990} In this work we have developed a theoretical model to treat normal Auger decay in a time-dependent framework, that is, the Auger decay proceeds at the same time that the pulse is interacting with the system and inducing the ionization. X-ray pulses are considered coherent at this timescale, and this has effects in the x-ray--matter interactions. By using this model we reveal two main effects in the Auger-electron--photoelectron coincidence caused by such short pulses, while those effects are hidden when the non-coincidence Auger spectrum is analyzed. The first effect is due to the time duration of the pulse. As the pulse length shortens, the Auger electron--photoelectron peaks broaden. For pulses much shorter than the core-hole lifetime, the lineshape of the peaks converge to a Lorentzian profile with a broadening given by the natural width of the core-hole state. We discuss the relation of this effect to the known bandwidth effects in resonant Auger. The second effect is due to the intensity of the pulse, related to the depletion of the ground state while the core-hole state is decaying by Auger processes. In this work, the x-ray pulse, although quite intense, is not intense enough to induce multiphoton effects, which have known effects on the atomic system. \cite{Young2010,Hoener2010,Schorb2012} The goal of the present study is to explore only the effects due to a fast, coherent control of the core-hole state rather than entering in a multiphoton regime in which the system is ionized several times. 

This manuscript is divided as follow: in Section \ref{sec:model} we describe the  theoretical model; in Section \ref{sec:results} we present our main results showing the effects on the electron-coincidence spectra due to the time duration as well as the intensity of the pulse; finally, is Section \ref{sec:conclusions} we present the conclusions and the outlook of our findings.


\section{Time-dependent model for Auger decay} \label{sec:model}

In this section we describe a time-dependent model to study coherent processes during Auger processes in an atomic system. The goal of this model is to allow us to treat the electron correlation that describe the Auger decay and the ionization of the x-rays in very short and comparable time scales. This model is similar to those used in previous works to study resonant Auger decay process. \cite{Nina2008,Picon2013_A,Picon2013_B} First, we assume that the dynamics of the system will be restricted to a small Hilbert space by using the ansatz
\begin{eqnarray} 
\vert \psi (t) \rangle &=& b_0 (t) \vert 0 \rangle + \sum_\varepsilon \sum_i b_{\varepsilon ; i}(t) \vert \varepsilon; i \rangle + \nonumber\\
&&\sum_{\varepsilon \varepsilon_a} \sum_{i j}  b_{\varepsilon \varepsilon_a ; i j}(t) \vert \varepsilon \varepsilon_a ; i j \rangle \; , \label{ansatz_Auger}
\end{eqnarray}
where $b_0$ stands for the amplitude of the ground state, $b_{\varepsilon ; i}$ for the core-excited states after x-ray ionization, and $b_{\varepsilon \varepsilon_a ; i j}$ for the final states after Auger decay. In order to construct the Hamiltonian of the system and derive the corresponding time-dependent Schr\"odinger equation, we need to choose a many-body theoretical framework for our basis. Although our time-dependent model can be extended to any framework, we limit ourselves to a Hartree-Fock based model.  

In the Hartree-Fock description the ground state is written as
\begin{eqnarray}
\vert 0 \rangle = \vert \{\alpha\} \rangle = \vert \{\Phi_a(1) \Phi_b(2) ... \Phi_n(N) \} \rangle
\end{eqnarray}
where the {curly brackets} refer to a Slater determinant, i.e. to all possible permutations (with the proper sign) among the spin orbitals $\Phi_\alpha$. In a second quantization notation, the one-electron and two-electron excitation can be written as
\begin{eqnarray}
&& \vert \varepsilon; i \rangle =  a^{\dagger}_\varepsilon a_i \vert \{\alpha\} \rangle \\
 && \vert \varepsilon \varepsilon_a; i j\rangle =  a^{\dagger}_\varepsilon a^{\dagger}_{\varepsilon_a} a_j a_i \vert \{\alpha\} \rangle
\end{eqnarray}

Hence, $b_{\varepsilon ; i}$ stands for excited states that differ from one orbital with respect to the ground state. In particular, the spin orbital $\Phi_i$ is {\{ replaced} by the continuum orbital $\Phi_\epsilon$. Similarly, the amplitudes of the final states after Auger decay are given by $b_{\varepsilon \varepsilon_a ; i j}$, in which two orbitals $\Phi_i$ and $\Phi_j$ are substituted by two continuum orbitals $\Phi_\epsilon$ and $\Phi_{\epsilon_a}$, representing the photoelectron and the Auger electron. In general, the notation $ij$ refers to bound states, while $\varepsilon \varepsilon_a$ refers to continuum states. The sums over continuum states contain both integrals of the energy variables and sums over discrete quantum numbers such as angular momenta and spin.

The Hamiltonian $\hat{H}$ of the atomic system can be written as the sum of the independent-particle effective Hamiltonian $\hat{H}_{\rm eff}$ plus a residual term $V_{a}$, which is a potential that quantifies the departure from the independent-particle picture. Considering also the coupling with the external electric field of the x-rays $\hat{V}_I$, the Schr\"odinger equation is given by
\begin{eqnarray} \label{SE_Auger}
i \frac{\partial}{\partial t} \vert \psi (t) \rangle = [\hat{H}_{\rm eff} + \hat{V}_{a} + \hat{V}_I (t)] \vert \psi (t) \rangle \; .
\end{eqnarray}

Including the ansatz given by Eq. (\ref{ansatz_Auger}) into the Schr\"odinger equation (\ref{SE_Auger}), we can derive the equations of motion (EOM) for the amplitudes $b_0,b_{\varepsilon ; i}$ and $b_{\varepsilon \varepsilon_a ; i j}$, see appendix \ref{sec:EOM}. The obtained EOMs are still difficult to solve from the numerical point of view. Within the so-called adiabatic approximation, sometimes also referred to as local approximation, \cite{Cederbaum1981,Domcke1991} those equations can be further reduced and be easily solved numerically. The adiabatic approximation is well-justified in Auger processes, in which the electron-correlation couplings responsible for the Auger decay do not significantly change within the wide range of the Auger electron energies. Within this approximation, and also the rotating-wave approximation, the dominant terms of the EOM are reduced to

\begin{eqnarray}  \nonumber 
i\, \dot{b}_0 (t) &=& E_0 \, b_0 (t)  -i \frac{\Gamma_I(t)}{2} b_0 (t)  \\
i\, \dot{b}_{\varepsilon ; i}(t)  &=& E_{\varepsilon ; i} {b}_{\varepsilon ; i}(t) -i \frac{\Gamma_{\varepsilon i,\varepsilon i}}{2} b_{\varepsilon ; i}(t) +  \nonumber \\  
&& \langle \varepsilon; i \vert V_I(t) \vert 0 \rangle \, b_0 (t)  \; , \nonumber \\
i\, \dot{b}_{\varepsilon \varepsilon_a ; i j} (t) &=&  E_{\varepsilon \varepsilon_a ; i j} \, b_{\varepsilon \varepsilon_a ; i j} (t) +  \nonumber \\
&& \langle  \varepsilon \varepsilon_a ; i j \vert V_{a}\vert  \varepsilon; i  \rangle \,  b_{\varepsilon ; i}(t) 
\;, \nonumber \\
\label{EOM_Auger_adb2}
\end{eqnarray}
where 
\begin{widetext}
\begin{eqnarray*}
\frac{\Gamma_{\varepsilon i,\varepsilon i}}{2} &=& \pi \sum_{\varepsilon'' \varepsilon'_a} \sum_{i'' j''} \langle  \varepsilon; i \vert V_{a}\vert  \varepsilon'' \varepsilon''_a ; i'' j''  \rangle \, 
\langle  \varepsilon'' \varepsilon''_a ; i'' j'' \vert V_{a}\vert  \varepsilon; i  \rangle  \;\; \delta(E_{\varepsilon'' \varepsilon''_a ; i'' j''} - E_{\varepsilon ; i}) \\
\frac{\Gamma_{I}(t)}{2} & = & \frac{\Omega^2(t)}{4} \sum_\varepsilon \sum_i  \langle 0 \vert \tilde{V}_I \vert \varepsilon; i \rangle  \langle \varepsilon; i \vert \tilde{V}_I \vert 0 \rangle 
\frac{\Gamma_{\varepsilon i,\varepsilon i}/ 2 }{ (E_{\varepsilon; i} -E_0 -\omega)^2 + \left({\Gamma_{\varepsilon i,\varepsilon i}/ 2}\right)^2 } \, \\
\end{eqnarray*}
\end{widetext}

The obtained EOM are easy to interpret. The energies of the ground state, core-excited states, and final states are given by $E_0$, $E_{\varepsilon; i}$, and $E_{\varepsilon \varepsilon_a ; i j}$ respectively. The Rabi frequency of the pulse is given by $\Omega(t)$, the frequency of the pulse by $\omega$, the dipole moments between ground state and core-excited states by $\langle 0 \vert \tilde{V}_I \vert \varepsilon; i \rangle$, where $\tilde{V}_I$ stands for the electric dipole moment $- \sum_j q_j\; {\bf r}_j \cdot {\bf s}$, where ${\bf s}$ is the polarization direction. The ionization rate of the ground state is related to the term $\Gamma_{I}(t)$, which depends on the envelope (intensity) of the pulse. The Auger transitions are given by the couplings $\langle  \varepsilon; i \vert V_{a}\vert  \varepsilon'' \varepsilon''_a ; i'' j''  \rangle$. The decay of the core-excited state is related to the term $\Gamma_{\varepsilon i,\varepsilon i}$, which is the sum of all different Auger transitions allowed in the system. 

The two-electron coincidence measurements, i.e. the measurement of the photoelectron and Auger electron in coincidence, is given by
\begin{eqnarray}
P(\varepsilon,\varepsilon_a) = \lim_{t\rightarrow \infty} \sum_{ij} \vert b_{{\varepsilon}{\varepsilon_a};ij} (t) \vert^2
\end{eqnarray}

In the previous formula, although is not written explicitly, we consider also the sum over the other quantum numbers of the photoelectron and Auger electron. The photoelectron spectrum and the Auger spectrum are given then by
\begin{eqnarray}
P_{ph}(\varepsilon)=\sum_{\varepsilon_a} P(\varepsilon,\varepsilon_a) \; \label{Ph_Spectrum}, \\
P_{a}(\varepsilon_a)=\sum_{\varepsilon} P(\varepsilon,\varepsilon_a) \; \label{Auger_Spectrum},
\end{eqnarray}
respectively. 

In general, equations (\ref{EOM_Auger_adb2}) do not have closed-analytical solutions and we need to solve numerically the coupled differential equations with Runge-Kutta, Crank-Nicolson or similar solver. However, there is a particular case, when the x-ray pulse is considered to have an ideal square envelope profile, that an analytical solution is easy to be derived. It is informative to assume the square pulse approximation to understand the origins of the main effects due to pulse duration and intensity. See Appendix \ref{sec:square_pulse} for more details. 

Assuming that the duration of the square pulse is given by $T$, the amplitude for the final states can be written as
\begin{widetext}
\begin{eqnarray}
\lim_{t\rightarrow \infty} \vert b_{{\varepsilon}{\varepsilon_a};ij} (t) \vert^2 =  \left| \Omega(t)\sum_{i'} \langle  \varepsilon \varepsilon_a ; i j \vert V_{a}\vert  \varepsilon; i'  \rangle  \langle \varepsilon; i' \vert \tilde{V}_I \vert 0 \rangle \frac{\sin[\left(\omega+E_0-E_{\varepsilon \varepsilon_a ; i j}-i\frac{\Gamma_{I}}{2}\right)T]}{\omega+E_0-E_{\varepsilon \varepsilon_a ; i j}-i\frac{\Gamma_{I}}{2}} \frac{1}{(E_{\varepsilon ; i'}-E_{\varepsilon \varepsilon_a ; i j})-i\frac{\Gamma_{\varepsilon i',\varepsilon i'}}{2}} \right|^2 . \nonumber \\
\label{Square_pulse}
\end{eqnarray}
\end{widetext}


The { factors} of (\ref{Square_pulse}) can be interpreted to gain insight to the physical process. The first two { factors} are related to the couplings of ionizing the ground state with the core-hole state decays into a particular final state. The third { factor} is a \textit{Sinc} function, related to the bandwidth of the pulse. The fourth { factor} is a Lorentzian function, related with the decay of the core-hole state. Because the pulse is chosen to have a temporal square profile, the Fourier transform has a Sinc shape profile. Note that the ionization rate of the ground state is included in this function. We use (\ref{Square_pulse}) to understand some of the effects we discuss in the next section. In particular, it is worth noting that if we consider that the ionization of the ground state is small and the pulse is very long then
the Sinc function becomes a Dirac-Delta function, then we recover the normal lineshape that we find in synchrotron experiments as we discuss in the following. The derived formula is similar to previous formulas for normal Auger spectra, see for example Refs. ~\cite{Vegh1994,Sheinerman2007,Penent2008} and references therein. The novelty of our approach is the explicit inclusion of the bandwidth effects and the depletion of the ground state described by the Sinc function.   

\section{Results}\label{sec:results}

In this section we present the effects on the photoelectron and the Auger electron spectra due to the pulse length and intensity of an x-ray pulse. We expect those effects to be significant when the pulse length of the pulse is comparable to the lifetime of the core-hole state, or when the intensity of the pulse is strong enough to induce depletion of the ground state in times that are also comparable with the core-hole lifetime.

\begin{figure}[t]
\begin{center}
\includegraphics[width=7.cm]{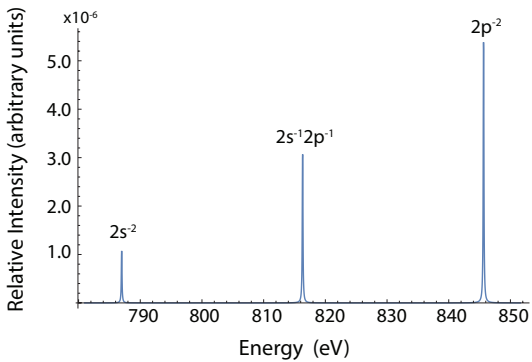}
\caption{Auger spectrum for a neon atom due to a 10 fs, $10^{13}$ W/cm$^2$ x-ray pulse. The three different peaks come from different Auger decay pathways into the final states 2s$^{-2}$, 2s$^{-1}$2p$^{-1}$, and 2p$^{-2}$.}
\label{Fig:auger}
\end{center}
\end{figure}

In our numerical simulations, solving the EOM given by (\ref{EOM_Auger_adb2}), we consider a Gaussian profile to model the x-ray pulse. For the sake of clarity and understanding the main effects, we will compare those numerical results with the analytical formula (\ref{Square_pulse}) for the square pulse. In our simulations we focus on the Ne atom. In order to calculate the necessary electric dipole and Auger transitions, we use a K-matrix approach to calculate the continuum orbitals. \cite{Starace1982,Demekhin2011} The bound states are expressed in the localized Pople's Gaussian basis set 6-311G. A Gaussian basis set is used so it will be possible to extend the present model to molecules by using a multi-center grid \cite{Demekhin2011}. The photon energy of the x-ray pulses is 1000 eV, well above the ionization threshold of the 1s orbital of Ne (experimentally it is 870.17 eV).\cite{Hitchcock1980} In our numerical simulations, the K-shell ionization threshold is found to be 892 eV, around 22 eV higher than the experimental value. This deviation results from neglecting orbital relaxation (if we consider relaxation, we obtain 869.9 eV). The ionization cross section at 1000 eV is calculated to be around 0.12 Mb. The core-hole lifetime is calculated to be 0.1192 eV, while the experimental value is around 0.27 eV. \cite{Coreno1999} This deviation is also due to relaxation; if we consider relaxation we obtain 0.27 eV. Those deviations are however not relevant for the time-dependent effects that we discuss in the next sections, which are reflected in the lineshape of the Auger electron -- photoelectron coincidence spectra and not in the position or height of the peaks. Solving the EOM (\ref{EOM_Auger_adb2}) and obtaining the amplitudes of the final states $b_{{\varepsilon}{\varepsilon_a};ij} (t)$ for times long after the x-ray absorption and core-hole decay, we obtain the Auger spectrum, using Eq. (\ref{Auger_Spectrum}), shown in Fig.\ref{Fig:auger}. The Auger spectra show three prominent peaks; the first peak involves transitions with the 2s electrons, the second peak involves transitions with the 2s and 2p electrons, and the third peak involves transitions with the 2p electrons. 

\begin{figure}[t]
\begin{center}
\includegraphics[width=6.5cm]{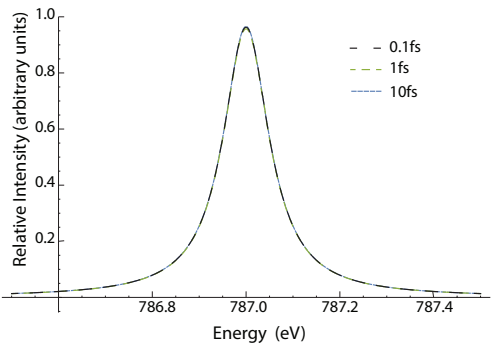}
\caption{Auger spectrum centered at 787 eV for three x-ray pulses with 0.1, 1 and 10 fs pulse length. The three Auger spectra overlap, and show no dependence on the pulse length. The intensity of the pulses is 10$^{15}$ W/cm$^2$. The K-shell ionization probability is small, less than $0.01\%$}
\label{Fig:bandwidth1}
\end{center}
\end{figure}

In the next section we discuss the effects on the photoelectron and Auger spectra when the pulse length of the pulses are comparable with the lifetimes of core-hole states. Then, in the following section, we discuss the effects that arise purely from the intensity of the x rays. 

\subsection{Pulse duration effects}

In this section we explore the effects on the electron spectra when the x-ray pulses have pulse lengths comparable to the lifetime of the core-hole state. We focus our discussion on the first Auger peak centered around 787 eV, with $2s^2$ final state, but the effects are also observed in the other peaks of the spectra.

\begin{figure}[t]
\begin{center}
\includegraphics[width=6.5cm]{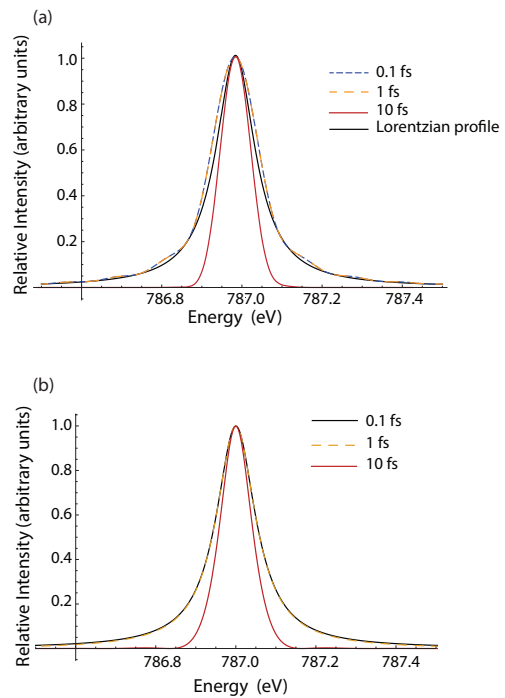}
\caption{(a) Auger electron-photoelectron coincidence spectrum centered at 787 eV for three x-ray pulses with 0.1, 1 and 10fs pulse length. A noticeable pulse duration effect is visible, as the spectrum broadens as pulse length is decreased. The intensity of the pulses are $10^{15}$ W/cm$^2$, $10^{14}$ W/cm$^2$ and $10^{13}$ W/cm$^2$, respectively. (b) Normalized probability obtained with formula (\ref{Square_pulse}) at the photoelectron energy 787 eV and using the corresponding pulse parameters to (a).}
\label{Fig:bandwidth2}
\end{center}
\end{figure}

We consider three pulse lengths that differ from the calculated 8.3 fs core-hole lifetime: 0.1 fs, 1 fs and 10 fs. We begin by considering low intensities in which the ground state is barely depleted (around 0.01\% excitation). For the three previous pulse lengths we consider $10^{15}$ W/cm$^2$, $10^{14}$ W/cm$^2$ and $10^{13}$ W/cm$^2$, respectively, to maintain the ionization of the ground state constant. In Fig. \ref{Fig:bandwidth1} we show the first peak of the Auger spectrum centered around 787 eV, and it shows no dependence on the pulse length. However, a dependence on pulse duration emerges when we consider the Auger electron-photoelectron coincidence spectrum. We select the photoelectron energy 208.5 eV with an energy resolution of 50 meV. Figure \ref{Fig:bandwidth2}(a) shows the Auger spectrum for the selected photoelectron energy that has a clear dependence on the duration of the pulse. For the sake of comparison, all the peaks shown have been properly normalized -- the heights of the peaks are taken to be one, in order to compare the obtained lineshape with different pulse parameters. First we note that for longer pulses, the broadening of the peak reduces. This is a bandwidth effect, similar to the bandwidth effect observed in resonant Auger processes \cite{Piancastelli2000}. This effect can be connected with nonresonant Auger Raman spectroscopy \cite{Viefhaus1998}, in which the photoelectron was detected in coincidence with the Auger electron in order to obtain Auger spectra with sub-lifetime energy widths. However, in our case, we note that the broadening of the peaks for the 0.1 fs and 1 fs pulses are similar, even though there is a factor 10 between the time duration of both pulses. In fact, the broadening of the peaks is not larger than the natural width of the $1s$ core hole. This effect is clear in the analytical formula (\ref{Square_pulse}) for a square pulse. The selected-photoelectron Auger coincidence spectrum is the result of the multiplication of the bandwidth of the pulse with the Lorentzian profile from the Auger decay. If the bandwidth is broad enough, i.e. with a pulse length much shorter than the core-hole lifetime, then the selected-photoelectron Auger lineshape is mainly given by the Lorentzian profile. In figure \ref{Fig:bandwidth2}(b) we show the lineshapes given by the formula (\ref{Square_pulse}) for the three different pulse lengths, showing the same trend as the selected-photoelectron Auger spectra of Fig. \ref{Fig:bandwidth2}(a). For longer pulses, the broadening of the pulse is mainly given by the bandwidth of the pulse, convoluted by the natural width of the core-hole state \cite{Viefhaus1998}. Note that \ref{Fig:bandwidth2}(a) slightly deviates from a perfect Lorentzian line-shape, the fine structure is due to the selection of the data in the energy range, without considering any smooth integration in the edge of the energy windows.  

\begin{figure}[h]
\begin{center}
\includegraphics[width=8.5cm]{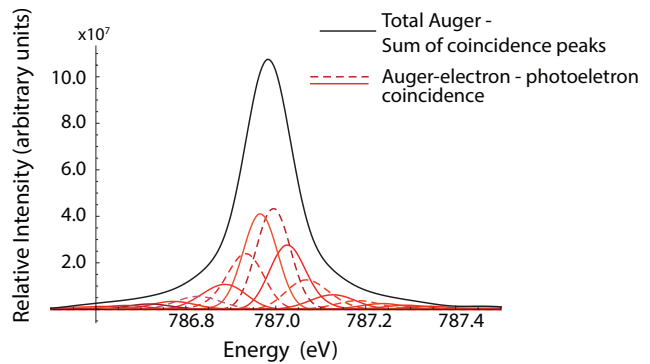}
\caption{Auger electron-photoelectron Auger spectra around the peak value of 208.5 eV. Their center depends on the selected photoelectron energy. By summing the coincidence peaks, we obtain the total Auger spectrum, indicated in the figure as the envelope with a perfect Lorentzian profile.}
\label{fig:envelope}
\end{center}
\end{figure}

We have also studied other Auger spectra with different photoelectron energies. In Fig. \ref{fig:envelope} we show different selected-photoelectron Auger spectra for the intermediate intensity with 10 fs pulse. Note that other selected-photoelectron Auger spectra also present similar lineshapes, deviated from a Lorentzian function, however, the sum of all of them (envelope line), has the expected Lorentzian profile. Note that by selecting the photoelectron energy, we also select the range of energies of the emitted Auger electron. This is explained by conservation of energy as in the case of resonant Auger processes.\cite{Piancastelli2000}

\subsection{Intensity effects}

In this section we focus on the effects on the photoelectron and Auger spectra that arise from the intensity of the pulse instead of the pulse duration. Similarly to the last section, we found no effects in the integrated Auger spectrum, but instead, by selecting the Auger electrons in coincidence with a particular photoelectron energy, we observe changes in the Auger electron-photoelectron lineshape.

\begin{figure}[h]
\begin{center}
\includegraphics[width=6.5cm]{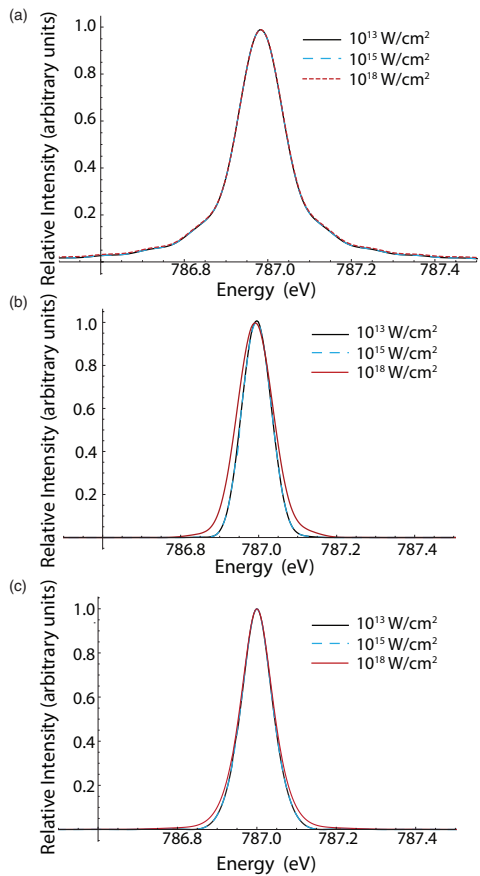}
\caption{Auger electron-photoelectron coincidence spectrum centered at 787 eV for three different intensities $10^{13}$, $10^{15}$, and $10^{18}$ W/cm$^2$ for a (a) 0.1-fs pulse and (b) 10-fs pulse. A noticeable intensity effect is visible only for the longer pulse. (c) Normalized probability obtained with formula (\ref{Square_pulse}) at the photoelectron energy 787 eV and using the same pulse parameters than (b).}
\label{fig:intensity1}
\end{center}
\end{figure}

\begin{figure}[h]
\begin{center}
\includegraphics[width=6.5cm]{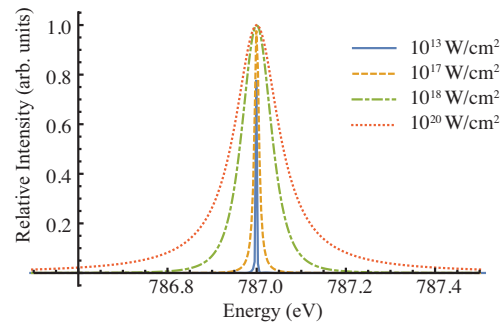}
\caption{Auger electron-photoelectron coincidence spectrum centered at 787 eV for four different intensities 10$^{13}$, 10$^{17}$, 10$^{18}$ and 10$^{20}$ W/cm$^2$ for a 500-fs pulse. Normalized probability obtained with formula (10) at the photoelectron energy 787 eV and using the same pulse parameters than Fig. \ref{fig:intensity1} (b).}
\label{fig:intensity2}
\end{center}
\end{figure}

We select the peak photoelectron energy, 208.5 eV, with an energy resolution of 50 meV and, we study three intensities; when the ground state is barely depleted 10$^{13}$ W/cm$^2$ (0.01\%), when the ground state is intermediately depleted 10$^{15}$ W/cm$^2$ (1\%), and when the ground state is completely depleted 10$^{18}$ W/cm$^2$ (100\%). We choose not to use higher intensities, so as to avoid entering the multiphoton regime that the model does not account for. In  Fig. \ref{fig:intensity1}(a) and Fig. \ref{fig:intensity1}(b) we show the selected-photoelectron Auger spectra for two different pulse lengths, 0.1 fs and 10 fs respectively. While the spectrum for the shorter pulse shows no intensity dependence, the spectrum for the longer pulse shows a clear dependence with the intensity. Specifically, the width of the peaks become broader as the intensity of the pulse is increased. Similar to the observed effect in the previous section, the broadening of the peak never becomes larger than the natural width of the core-hole state. Hence, we observe that the strong population of the core-hole state in times comparable with the core-hole lifetime modifies the lineshape of the spectrum; however, this is not noticeable for the 0.1-fs pulse length. {We observe that this effect persists at longer 500-fs pulses, see figure \ref{fig:intensity2} (no multiphoton processes are considered). Hence, this effect exists in the continuous-wave limit but gradually disappears as the pulses become shorter.}

In order to understand the intensity dependence, and the connection between the spectrum and the broadening caused by the intensity of the pulse and the pulse length, we resort to the analytical formula (\ref{Square_pulse}) for the square pulse to gain additional insight. In  Fig. \ref{fig:intensity1}(c) we show the lineshape profile calculated with the analytical formula for the three different intensities for the 10-fs pulse, showing a similar trend than the selected-photoelectron Auger spectra. The intensity dependence is contained in the Sinc function (\ref{Square_pulse}) via the ionization rate given by $\Gamma_I$. If we increase the ionization rate, the Sinc function begins to lose the oscillations and the tail starts to increase, and consequently the central peak becomes broader. However, if the Sinc function becomes very broad when the intensity is very high, then the lineshape is dominated by the Lorentzian function from the core-hole decay. This explains why we do not observe any intensity effect for the 0.1-fs pulse, as the broadening of the peak is already dominated by the Lorentzian function and the intensity does not play a role in the lineshape of the spectrum.

\section{Conclusions} \label{sec:conclusions}

We have studied the effects in the Auger electron -photoelectron coincidence spectra due to ultrashort x-ray pulses. These pulses modify the core-hole population within the timescale of the core-hole lifetime. By using a time-dependent model for Ne atom, we are able to explain two effects that arise from the pulse length and intensity of the pulse. These effects are observable in a photoelectron-Auger electron coincidence measurement. When the pulse is shorter, we observe a broadening of the Auger electron--photoelectron peaks. This effect in the lineshape is similar to the known bandwidth broadening in resonant Auger. When the pulse is intense enough to deplete the ground state, and populate the core-hole states during the Auger lifetime, we  observe a broadening of the Auger electron--photoelectron peaks. However, in both effects the broadening of the peaks is always limited by the natural width of the core-hole state. When the total Auger electron spectrum is analyzed, by integrating over all photoelectron energies, the normal Auger lineshape is recovered, exhibiting no dependence on the time duration or intensity of the pulse.

The results shown in this work provide better understanding of the x-ray interactions with pulses that are comparable to the lifetime of the core-hole states. These effects are expected to be present in any other physical system, such as molecules, clusters, and nanoparticles. The theoretical framework presented here is ideal to study Auger processes of future ultrafast experiments with femtosecond and attosecond x-ray pulses.




\begin{acknowledgments}
A.P. acknowledge fruitful discussions with A. Lopez-Bezanilla and S.H. Southworth. We gratefully acknowledge the computing resources provided on Blues, a high-performance computing cluster operated by the Laboratory Computing Resource Center at Argonne National Laboratory. This work was supported in part by the U.S. Department of Energy, Office of Science, Office of Workforce Development for Teachers and Scientists (WDTS) under the Science Undergraduate Laboratory Internship (SULI) program. This material is based upon work supported by the U.S. Department of Energy, Office of Science, Basic Energy Sciences, Chemical Sciences, Geosciences, and Biosciences Division and supported the Argonne group under contract no. DE-AC02-06CH11357.
This research used resources of the Argonne Leadership Computing Facility, which is a DOE Office of Science User Facility supported under Contract DE-AC02-06CH11357.
\end{acknowledgments}

\appendix

\section{Equations of motion} \label{sec:EOM}
When we include the ansatz (\ref{ansatz_Auger}) into the time-dependent Schr\"odinger equation, we obtain the following equations for the amplitudes
\begin{eqnarray} 
i\, \dot{b}_0 (t) &=& E_0 \, b_0 (t) + \sum_\varepsilon \sum_i  \langle 0 \vert V_I(t) \vert \varepsilon; i \rangle \, b_{\varepsilon ; i}(t)  \; , \label{EOM_Auger_RPA_1} \\
i\, \dot{b}_{\varepsilon ; i}(t) &=& E_{\varepsilon ; i} {b}_{\varepsilon ; i}(t) + \langle \varepsilon; i \vert V_I(t) \vert 0 \rangle \, b_0 (t) + \nonumber \\
&& \sum_{\varepsilon' \neq \varepsilon} \sum_{i'\neq i} \langle  \varepsilon; i \vert V_{a}\vert \varepsilon'; i'  \rangle \,  b_{\varepsilon'; i' }(t) +  \nonumber \\
&& \sum_{\varepsilon' \varepsilon_a} \sum_{i' j'} \langle  \varepsilon; i \vert V_{a}\vert  \varepsilon' \varepsilon'_a ; i' j'  \rangle \,  b_{\varepsilon' \varepsilon'_a ; i' j'}(t) \; , \label{EOM_Auger_RPA_2}\\
i\, \dot{b}_{\varepsilon \varepsilon_a ; i j} (t) &=&  E_{\varepsilon \varepsilon_a ; i j} \, b_{\varepsilon \varepsilon_a ; i j} (t) +  \nonumber \\
&& \sum_{\varepsilon'} \sum_{i'} \langle  \varepsilon \varepsilon_a ; i j \vert V_{a}\vert  \varepsilon'; i'  \rangle \,  b_{\varepsilon' ; i'}(t) + \nonumber \\
&& \sum_{\varepsilon' \varepsilon'_a \neq \varepsilon \varepsilon_a } \sum_{i' j' \neq i j} \langle  \varepsilon \varepsilon_a ; i j \vert V_{a}\vert  \varepsilon' \varepsilon_a' ; i' j'  \rangle \,  b_{\varepsilon' \varepsilon'_a ; i' j'}(t)
\;, \nonumber \\
\label{EOM_Auger_RPA_3}
\end{eqnarray}
We have neglected the terms 
\begin{eqnarray*}
&&\langle \varepsilon; i \vert  \hat{V}_{a} \vert 0 \rangle \approx 0 \\
&&\langle \varepsilon \varepsilon_a ; i j \vert  \hat{V}_{a} \vert 0 \rangle \approx 0
\end{eqnarray*}
In any case, because of the Brillouin theorem, the first product is identical to zero in the Hartree-Fock approximation.

In inner-shell ionization, when the ionization may come from several degenerate states or close by in energies, for example the ionization of a 3d electron in Xe, then the Random Phase Approximation (RPA) provides a good theoretical description of the involved electron correlations, see for example Ref. \cite{Becker_book}. The RPA can also be applied in the calculations for the Auger decay transitions. In this work, we consider those electron-correlation couplings to be zero, that is
\begin{eqnarray*}
&& \langle  \varepsilon; i \vert V_{a}\vert \varepsilon'; i'  \rangle \approx 0 \\
&& \langle  \varepsilon \varepsilon_a ; i j \vert V_{a}\vert  \varepsilon' \varepsilon_a' ; i' j'  \rangle  \approx 0
\end{eqnarray*}

The system of equations (\ref{EOM_Auger_RPA_1}),(\ref{EOM_Auger_RPA_2}), and (\ref{EOM_Auger_RPA_3}) can be further decoupled by using the adiabatic approximation, also called local approximation \cite{Cederbaum1981,Domcke1991}. For the ionization part, also use the adiabatic approximation, also known in the Quantum Optics literature as Markov approximation, see for example \cite{Knight1990,Tannoudji1977}. Within these approximations, the EOMs can be reduced with the derivation of decay rates $\Gamma$ that accounts for the ionization of the ground states and the Auger transitions of the core-excited state, and Stark shifts $R$ that accounts for the dephasing introduced by the continuum part that has been decoupled:
\begin{widetext}
\begin{eqnarray}  \nonumber
&& i\, \dot{b}_0 (t) = E_0 \, b_0 (t)  -i \left[ {\Gamma_I(t) \over 2} + i R_I (t) \right] b_0 (t)  \\
&& - \sum_\varepsilon \sum_i  \langle 0 \vert V_I \vert \varepsilon;i \rangle \int_{t_0}^t dt'  \sum_{\varepsilon'\neq\varepsilon} \sum_{i'\neq i} \left[ {\Gamma_{\varepsilon i,\varepsilon' i'} \over 2} + i R_{\varepsilon i,\varepsilon' i'} \right] b_{\varepsilon' ; i'}(t')   \; e^{-i (E_{\varepsilon ; i} + R_{\varepsilon i, \varepsilon i} - i \Gamma_{\varepsilon i, \varepsilon i}/2) (t-t')}   \; , \nonumber \\
&& i\, \dot{b}_{\varepsilon ; i}(t)  = E_{\varepsilon ; i} {b}_{\varepsilon ; i}(t) + \langle \varepsilon; i \vert V_I(t) \vert 0 \rangle \, b_0 (t)   
-i \sum_{\varepsilon'} \sum_{i'} \left[ {\Gamma_{\varepsilon i,\varepsilon' i'} \over 2} + i R_{\varepsilon i,\varepsilon' i'} \right] b_{\varepsilon' ; i'}(t)  \; , \nonumber \\
&& i\, \dot{b}_{\varepsilon \varepsilon_a ; i j} (t) =  E_{\varepsilon \varepsilon_a ; i j} \, b_{\varepsilon \varepsilon_a ; i j} (t) +  
\sum_{\varepsilon'} \sum_{i'} \langle  \varepsilon \varepsilon_a ; i j \vert V_{a}\vert  \varepsilon'; i'  \rangle \,  b_{\varepsilon' ; i'}(t) 
\;, \nonumber \\
\label{EOM_Auger_adb}
\end{eqnarray}

where

\begin{eqnarray*}
{\Gamma_{\varepsilon i,\varepsilon' i'} \over 2} &=& \pi \sum_{\varepsilon'' \varepsilon'_a} \sum_{i'' j''} \langle  \varepsilon; i \vert V_{a}\vert  \varepsilon'' \varepsilon''_a ; i'' j''  \rangle \, 
\langle  \varepsilon'' \varepsilon''_a ; i'' j'' \vert V_{a}\vert  \varepsilon'; i'  \rangle  \;\; \delta(E_{\varepsilon'' \varepsilon''_a ; i'' j''} - E_{\varepsilon' ; i'}) \\
R_{\varepsilon i,\varepsilon' i'} & = & -i \sum_{i'' j''} \langle  \varepsilon; i \vert V_{a}\vert  \varepsilon'' \varepsilon''_a ; i'' j''  \rangle \, 
\langle  \varepsilon'' \varepsilon''_a ; i'' j'' \vert V_{a}\vert  \varepsilon'; i'  \rangle \;\; P \left[ {1 \over E_{\varepsilon'' \varepsilon''_a ; i'' j''} - E_{\varepsilon' ; i'}}\right] \\
{\Gamma_{I}(t) \over 2} & = & {\Omega^2(t)\over4} \sum_\varepsilon \sum_i  \langle 0 \vert \tilde{V}_I \vert \varepsilon; i \rangle  \langle \varepsilon; i \vert \tilde{V}_I \vert 0 \rangle 
{\Gamma_{\varepsilon i,\varepsilon i}/ 2 \over (E_{\varepsilon; i} +R_{\varepsilon i,\varepsilon i} -E_0 -\omega)^2 + \left({\Gamma_{\varepsilon i,\varepsilon i}/ 2}\right)^2 } \, \\
R_{I}(t)  & = & - {\Omega^2(t)\over4} \sum_\varepsilon \sum_i  \langle 0 \vert \tilde{V}_I \vert \varepsilon; i \rangle  \langle \varepsilon; i \vert \tilde{V}_I \vert 0 \rangle 
{E_{\varepsilon; i} +R_{\varepsilon i,\varepsilon i} -E_0 -\omega \over (E_{\varepsilon; i} +R_{\varepsilon i,\varepsilon i} -E_0 -\omega)^2 + \left({\Gamma_{\varepsilon i,\varepsilon i}/ 2}\right)^2 }
\end{eqnarray*}
\end{widetext}

The symbol $P$ stands for the principal value. By taking only the dominant contributions, the Stark shifts are smaller than the decay rates, we obtain the EOM given in (\ref{EOM_Auger_adb2}).

\section{Auger decay for a square pulse} \label{sec:square_pulse}

In this section we show the derivation of the analytical formula (\ref{Square_pulse}) for the special case in which the electric field is an ideal square pulse. First, we calculate the line shape profile of the photoelectron, by using the second equation of the EOM (\ref{EOM_Auger_adb}), and then we can calculate the amplitudes giving rise to the Auger electron spectrum.

Considering a square pulse from $t_0$ to $t_1$, the photoelectron amplitude for a time $t$ that satisfies $t_0<t<t_1$, using the second line of Eqs. (\ref{EOM_Auger_adb}),
\begin{widetext}
\begin{eqnarray}
b_{\varepsilon,i}(t) =-i \int_{t_0}^{t}\!\! dt'  \langle \varepsilon; i \vert \tilde{V}_I \vert 0 \rangle  \; \epsilon (t') \left[e^{-iE_0(t'-t_0)} e^{-{\Gamma_{I}\over 2}(t'-t_0)} \right] e^{-i(E_{\varepsilon ; i} - i{\Gamma_{\varepsilon i,\varepsilon i}\over 2})(t-t')} \; ,
\label{Photoe2_Sq}
\end{eqnarray}

where $\langle \varepsilon; i \vert V_I(t) \vert 0 \rangle =\langle \varepsilon; i \vert \tilde{V}_I \vert 0 \rangle  \; \epsilon (t)$, and $\epsilon (t)$ is the electric field that can be written as $\epsilon (t)=\Omega(t) \cos(\omega t)$. If we perform the integration, we obtain:
\begin{eqnarray}
b_{\varepsilon,i}(t) = \frac{\Omega(t)}{2}\langle \varepsilon; i \vert \tilde{V}_I \vert 0 \rangle e^{-i(\omega+E_0)(t-t_0)} \left[
{1-e^{i(\omega+E_0-E_{\varepsilon ; i})(t-t_0)} e^{-({\Gamma_{\varepsilon i,\varepsilon i}\over 2}-{\Gamma_{I}\over 2})(t-t_0)} \over (\omega+E_0-E_{\varepsilon ; i}) + i({\Gamma_{\varepsilon i,\varepsilon i}\over 2}-{\Gamma_I\over 2})}
\right] e^{-{\Gamma_{I}\over 2}(t-t_0)} \; .
\label{Photoe2_Sq_2}
\end{eqnarray}
Similarly, for times $t$ that $t>t_1$, the solution is given by:
\begin{eqnarray}
b_{\varepsilon,i}(t) &=& b_{\varepsilon;i}(t_1) e^{-iE_{\varepsilon ; i}(t-t_1)} e^{-{\Gamma_{\varepsilon i,\varepsilon i}\over 2}(t-t_1)} \nonumber \\ 
&=& \frac{\Omega(t)}{2}\langle \varepsilon; i \vert \tilde{V}_I \vert 0 \rangle  e^{-i(\omega+E_0)(t_1-t_0)} e^{-iE_{\varepsilon ; i}(t-t_1)} \left[
{1-e^{i(\omega+E_0-E_{\varepsilon ; i})(t_1-t_0)} e^{-({\Gamma_{\varepsilon i,\varepsilon i}\over 2}-{\Gamma_{I}\over 2})(t_1-t_0)} \over (\omega+E_0-E_{\varepsilon ; i}) + i({\Gamma_{\varepsilon i,\varepsilon i}\over 2}-{\Gamma_I\over 2})}
\right] e^{-{\Gamma_{I}\over 2}(t_1-t_0)} e^{-{\Gamma_{\varepsilon i,\varepsilon i}\over 2}(t-t_1)} \; .
\label{Photoe2_Sq_3}
\end{eqnarray}
\end{widetext}
The time evolution of the Auger electron amplitude is given by the third line of Eqs. (\ref{EOM_Auger_adb}). Similarly to the amplitude of the core-excited state, we can transform the differential equation in its integral form. Then, considering a square pulse from $t_0$ to $t_1$, we can split the integral in two terms, one term when the time evolution of the system is interacting with the square pulse, and another accounting for the free propagation of the system after the square pulse:
\begin{widetext}
\begin{eqnarray}
{b}_{\varepsilon \varepsilon_a ; i j} (t) = &&-i \sum_{i'} \! \langle  \varepsilon \varepsilon_a ; i j \vert V_{a}\vert  \varepsilon; i'  \rangle \int_{t_0}^{t_1}dt' b_{\varepsilon;i'}(t') e^{-i E_{\varepsilon \varepsilon_a ; i j} (t-t')} \nonumber \\
 && -i \sum_{i'}\! \langle  \varepsilon \varepsilon_a ; i j \vert V_{a}\vert  \varepsilon; i'  \rangle \int_{t_1}^{t}dt' b_{\varepsilon;i'}(t') e^{-i E_{\varepsilon \varepsilon_a ; i j} (t-t')} \; ,
\label{Auger_ampl1_Sq}
\end{eqnarray}
for $t>t_1$. The solutions for $b_{\varepsilon;i} (t)$ are given by Eqs. (\ref{Photoe2_Sq_2}) and (\ref{Photoe2_Sq_3}). If we perform the corresponding integrations, we obtain
\begin{eqnarray}
&& {b}_{\varepsilon \varepsilon_a ; i j} (t) = - \sum_{i'} \! \langle  \varepsilon \varepsilon_a ; i j \vert V_{a}\vert  \varepsilon; i'  \rangle  \frac{\Omega(t)}{2}\langle \varepsilon; i' \vert \tilde{V}_I \vert 0 \rangle  \hspace{10cm} \nonumber \\ 
&& \; \left\{  \left[
{1 \over [(\omega+E_0-E_{\varepsilon \varepsilon_a ; i j})-i{\Gamma_{I}\over2}]}     {1 \over [(E_{\varepsilon;i'}-E_{\varepsilon \varepsilon_a ; i j})-i{\Gamma_{\varepsilon i',\varepsilon i'}\over2}]}
\right] \right. \nonumber \\
&& \hspace{2cm} \left[ -1 + e^{-[{\Gamma_I\over2}+i(\omega+E_0-E_{\varepsilon \varepsilon_a ; i j})](t_1-t_0)} 
\right] e^{-i E_{\varepsilon \varepsilon_a ; i j} (t-t_0)}   +  \nonumber \\
&& \;\;  \left[
{1 \over [(\omega+E_0-E_{\varepsilon;i'}) + i({\Gamma_{\varepsilon i',\varepsilon i'}\over 2}-{\Gamma_{I}\over 2})]}     {1 \over [(E_{\varepsilon;i'}-E_{\varepsilon \varepsilon_a ; i j})-i{\Gamma_{\varepsilon i',\varepsilon i'}\over2}]}
\right] \nonumber \\
&& \hspace{2cm} \; \left. \left[ -1 + e^{-[({\Gamma_{I}\over2}-{\Gamma_{\varepsilon i',\varepsilon i'}\over2})+i(\omega+E_0-E_{\varepsilon;i'})](t_1-t_0)} 
\right] e^{-[{\Gamma_{\varepsilon i',\varepsilon i'}\over2}+i E_{\varepsilon;i'}](t-t_0)}\right\} \; , \nonumber \\
\label{Auger_ampl1_Sq_2}
\end{eqnarray}
\end{widetext}

At much longer times than the core-hole lifetime, all the population in the core-hole states has already decayed into the final states. Hence, the Auger photoelectron spectra will be given by the modulus square of the final states amplitude, considering long times until the electrons reach the detectors. For long times, please note that the second line of equation (\ref{Auger_ampl1_Sq_2}) tends to zero.



\begin{references}
%
\bibitem{Allaria2010} E. Allaria {\em et al.}, New J. Phys. {\bf 12}, 075002 (2010).
%
\bibitem{Amann2012} J. Amann {\em et al.}, Nat. Photon. {\bf 6}, 693 (2012).
%
\bibitem{Hartmann2014} N. Hartmann {\em et al.}, Nat. Photon. {\bf 8}, 706 (2014).
%
\bibitem{Ratner2015} D. Ratner {\em et al.}, Phys. Rev. Lett. {\bf 114}, 054801 (2015).
%
\bibitem{Popmintchev2012} T. Popmintchev {\em et al.}, Science {\bf 336}, 1287 (2012).
%
\bibitem{Silva2015} F. Silva, S.M. Teichmann, S.L. Cousin, and J. Biegert, Nat. Commun. {\bf 6}, 6611 (2015).
%
\bibitem{Bostedt2016} C. Bostedt {\em et al.}, Rev. Mod. Phys. {\bf 88}, 015007 (2016).

\bibitem{Marinelli2015} A. Marinelli, D. Ratner, A. Lutman, J. Turner, J. Welch, F.-J. Decker, H. Loos, C. Behrens, S. Gilevich, A. Miah- nahri, {\em et al.} Nat. Commun. 6 (2015). 

\bibitem{Lutman2013} A. Lutman, R. Coffee, Y. Ding, Z. Huang, J. Krzywinski, T. Maxwell, M. Messerschmidt, and H.-D. Nuhn, Phys. Rev. Lett. 110, 134801 (2013).
%
\bibitem{Schmaltz2015} C.E. Liekhus-Schmaltz {\em et al.}, Nat. Commun. {\bf 6}, 8199 (2015).
%
\bibitem{Picon2016} A. Pic\'on {\em et al.}, Nature Commun. {\bf 7}, 11652 (2016).
%
\bibitem{Lehmann2016} C.S. Lehmann {\em et al.}, Phys. Rev. A {\bf 94}, 013426 (2016).
%
\bibitem{Ken2016} Ken R. Ferguson {\em et al.}, Science Advances {\bf 2}, e1500837 (2016).


\bibitem{Nina2008} Nina Rohringer and Robin Santra, Phys. Rev. A {\bf 77}, 053404 (2008).
%
\bibitem{Kanter2011} E.P. Kanter {et al.}, Phys. Rev. Lett. {\bf 107}, 233001 (2011).
%
\bibitem{JiCai2010} Ji-Cai Liu, Yu-Ping Sun, Chuan-Kui Wang, Hans {\r A}gren, and Faris Gel'mukhanov, Phys. Rev. A {\bf 81}, 043412 (2010).
%
\bibitem{Nikolopoulos2011} L.A.A. Nikolopoulos, T.J. Kelly, and J.T. Costello, Phys. Rev. A {\bf 84}, 063419 (2011).
%
\bibitem{Demekhin2013} Ph.V. Demekhin and L.S. Cederbaum, J. Phys. B: At. Mol. Opt. Phys. {\bf 46}, 164008 (2013).
%
\bibitem{Muller2015} D. M\"uller and Ph.V. Demekhin, J. Phys. B: At. Mol. Opt. Phys. {\bf 48}, 075602 (2015).
\bibitem{Mazza2012} T Mazza, K G Papamihail, P Radcliffe, W B Li, T J Kelly, J T Costello, S Düsterer, P Lambropoulos, and M Meyer, J. Phys. B: At. Mol. Opt. Phys. {\bf 45}, 141001 (2012).
%
\bibitem{Picon2013_A} A. Pic\'on, C. Buth, G. Doumy, B. Kr\"assig, L. Young, and S.H. Southworth, Phys. Rev. A {\bf 87}, 013432 (2013).
%
\bibitem{Picon2013_B} Antonio Pic\'on, Phay J. Ho, Gilles Doumy, and Stephen H Southworth, New J. Phys. {\bf 15}, 083057 (2013).
%
\bibitem{Bernhard2013} Bernhard W. Adams {\em et al.}, Journal of Modern Optics {\bf 60}, 2 (2013)
%
\bibitem{Chatterjee2015} Souvik Chatterjee and Takashi Nakajima, Phys. Rev. A {\bf 91}, 043413 (2015).
\bibitem{Haxton2014} D.J. Haxton and C.W. McCurdy, Phys. Rev. A {\bf 90}, 053426 (2014).
%
\bibitem{Picon2015} Antonio Pic\'on, Jordi Mompart, and Stephen H. Southworth, New J. Phys. {\bf 17}, 083038 (2015).

\bibitem{Mehlhorn1990} W. Mehlhorn, AIP Conf. Proc. {\bf 215}, 465 (1990)
\bibitem{Young2010}
L. Young {\em et al.}, Nature {\bf 466}, 56 (2010).
%
\bibitem{Hoener2010}
M. Hoener {\em et al.}, Phys. Rev. Lett. {\bf 104}, 253002 (2010).
%
\bibitem{Schorb2012}
S. Schorb {\em et al.}, Phys. Rev. Lett. {\bf 108}, 233401 (2012).
%
\bibitem{Cederbaum1981} L.S. Cederbaum and W. Domcke, J. Phys. B {\bf 14}, 4665 (1981). 
%
\bibitem{Domcke1991} W. Domcke, Phys. Rep. {\bf 208}, 97 (1991).
%
\bibitem{Becker_book} M.Ya. Amusia chapter 1: {\em Theory of photoionization: VUV and soft x-ray frequency region}. Editors U. Becker and D.A. Shirley, {\em VUV and soft X-ray photoionization}. New York: Plenum Press (1996).
%
\bibitem{Tannoudji1977} Cohen-Tannoudji, R. Balian, S. Haroche, S. Liberman (Eds.), {\em Frontiers of Laser Spectroscopy}, North-Holland, Amsterdam (1977).
%
\bibitem{Knight1990} P.L. Knight, M.A. Lauder, and B.J. Dalton, Phys. Rep. {\bf 190}, 1 (1990).
%
\bibitem{Vegh1994} L. V{\'e}gh and J.H. Macek, Phys. Rev. A {\bf 50}, 4031 (1994).
%
\bibitem{Sheinerman2007} S. Sheinerman, P. Lablanquie, and F. Penent, J. Phys. B: At. Mol. Opt. Phys. {\bf 40}, 1889 (2007).
%
\bibitem{Penent2008} F. Penent {\em et al.}, J. Phys. B: At. Mol. Opt. Phys. {\bf 41}, 045002 (2008).
%
\bibitem{Starace1982} A.F. Starace, {\em Theory of Atomic Photoionization}, Handbuch der Physik (Springer, Berlin, 1982), Vol. 31.
%
\bibitem{Demekhin2011} Ph.V. Demekhin, A. Ehresmann, and V.L. Sukhorukov, J. Chem. Phys. {\bf 134}, 024113 (2011).
%
\bibitem{Hitchcock1980} A.P. Hitchcock and C.F. Brion, J. Phys. B {\bf 13}, 3269 (1980).
%
\bibitem{Coreno1999} M. Coreno, {\em et al.}, Phys. Rev. A {\bf 59}, 2494 (1999).
%
\bibitem{Piancastelli2000} M.N. Piancastelli, J. Electron Spectrosc. Relat. Phemon. {\bf 107}, 1 (2000).
%
\bibitem{Viefhaus1998} J. Viefhaus {\em et al.}, Phys. Rev. Lett. {\bf 80}, 1618 (1998).
\end{references}

\end{document}